\def\beq{\begin{equation}}
\def\eeq{\end{equation}}

\def\hbar{\mathchar'26\mkern-9muh}

\def\MC{Monte Carlo}

\def\vMC{variational \MC}

\def\VMC{Variational \MC}

\def\dMC{diffusion \MC}

\def\DMC{Diffusion \MC}

\def\r2a{{\bf r}^{\bf a}}

\documentstyle[aps,floats]{revtex}
\reversemarginpar

\begin{document}
\title{Monte-Carlo studies of bosonic van der Waals clusters.}
\author{M. Meierovich, A. Mushinski and M.P. Nightingale }
\address{
  Department of Physics,\\
  University of Rhode Island,\\
  Kingston, RI 02881.
}
\maketitle

\begin{abstract}

In a previous paper \cite{andrei}, we developed a form of variational
trial wave function and applied it to van der Waals clusters:  five or
less atoms of Ar and Ne modeled by the Lennard-Jones potential.  In
addition, we tested the trial functions for a hypothetical, light atom
resembling Ne but with only half its mass.  We did not study atoms such
as He$^{4}$ with larger de Boer parameters, i.e., systems in which the
zero point energy plays a more important role relative to the potential
energy.  This is the main purpose of the present paper.  In fact, we
study clusters to the very limit where the zero-point energy destroys
the ground state as a bound state.  A simple picture of this un-binding
transition predicts the power law with which the energy vanishes as the
de Boer parameter approaches its critical value and the power of the
divergence of the the size of the clusters in this limit.  Our
numerical results are in agreement with these predictions.

\end{abstract}

\section{Introduction}

We consider clusters of bosonic Lennard-Jones atoms for which we present
two sets of results. Firstly, there are improved estimates of the
ground state energies of
systems studied previously with variational Monte Carlo \cite{andrei}.
The improvements
were obtained with a modified diffusion Monte Carlo algorithm \cite{we},
similar
to Ref. \onlinecite{umrigar}.
Secondly, we study the behavior of the clusters for small masses. That is,
in reduced units such that the Lennard-Jones pair potential has the form
$r^{-12}-2 r^{-6}$, the only independent parameter in the Schr\"odinger
equation is the reduced inverse mass $m^{-1}$,
a quantity proportional to the square of the de Boer parameter.
As the de Boer parameter increases, the relative importance of the
zero-point energy and the ground state energy of a cluster increase,
as does its size.  At a critical value of the de Boer parameter the ground
state energy $E_0$ vanishes while average cluster size $\langle r \rangle$,
as defined below, diverges and the cluster ceases
to exist in a bound state.

For the simple case of a dimer one can show that
\begin{eqnarray}
E_0~|_{m \downarrow m_{\rm c}} & \sim & \Delta m^2, \nonumber \\
\langle r \rangle ~|_{m \downarrow m_{\rm c}} & \sim & \Delta m^{-1}.
\label{eq.critical}
\end{eqnarray}
where $\Delta m = m-m_{\rm c}$ with $m_{\rm c}$ the critical value of the
reduced mass.  Note that $m_{\rm c}$ depends on $N$, the number of atoms
in the cluster and is expected to be a monotically decreasing function of
$N$.
The mathematical mechanism that yields Eqs.~(\ref{eq.critical}) is
the following.
Two scattering states forming a complex conjugate pair merge at zero
momentum to produce two states with
``complex momentum": a physically acceptable bound state and a state with
unacceptable behavior at infinity.  This mechanism is probably not limited
to the dimer and
it is quite plausible that Eqs.~(\ref{eq.critical}) apply in general
to clusters of any finite size.

\section{Results}
\label{Results}

Table ~\ref{tab.compare} shows the comparison between the ground state
energy estimates obtained by using \vMC~\cite{andrei} and our improved
\dMC\ algorithm.  Results obtained by \vMC\ suffer from a systematic
bias, i.e., if we denote by $E_{\rm T}=\langle \psi_{\rm T}|{\cal
H}|\psi_{\rm T} \rangle$, the variational estimate obtained a given
normalized trial state $|\psi_{\rm T} \rangle$, one has $E_0 < E_{\rm
T}$, where the equality holds only if the trial function is the exact
ground state wave function.

If one defines
\beq
\chi^2=\langle \psi_{\rm T}|({\cal H}-E_0)^2|\psi_{\rm T} \rangle
\eeq
the following inequality holds (see Ref. \onlinecite{andrei} for details and
references):
\beq
0<E_{\rm T}-E_0<{\chi^2\over E_1-E_0},
\label{eq.bound}
\eeq
where $E_1$ is the energy of the first, totally symmetric excited state.
To estimate the number of correct digits in the variational estimate of the
ground state and to ascertain how good a bound inequality~(\ref{eq.bound})
is we introduce the following quantities:
\begin{eqnarray}
Q'  &=& -\log_{10} \frac{\chi^2}{(E_1 - E_{\rm T})|E_{\rm T}|} \\ \nonumber
Q'' &=& \frac{\chi^2}{(E_1 - E_{\rm T})(E_{\rm T} - E_0)}.
\label{eq.bias}
\end{eqnarray}
The results are shown in Table~\ref{tab.compare}.  Quite remarkably,
the bound given in Eq.~(\ref{eq.bound}) is very tight.

\begin{table}
\begin{tabular} {rcdddd}
\multicolumn{1}{c}{}&
\multicolumn{1}{c}{$N$} &
\multicolumn{1}{c}{$E_{\rm T}$} &
\multicolumn{1}{c}{$E_0$} &
\multicolumn{1}{c}{$Q'$} &
\multicolumn{1}{c}{$Q''$} \\
\tableline

%==========================
%Atom		Number of atoms 	Upper bound Energy(Andrei's VMC)	Energy(My DMC)
%%Q'(Bias)	Q''(Relative bias)
%==========================
Ar			&3			&-2.553335364(1)		&-2.553335375(2)	&11.9		&---		\\
Ne			&			&-1.7195589(3)			&-1.7195586(5)		&7.40		&---		\\
{\tiny $1\over 2$}-Ne	&			&-1.308443(2)			&-1.308444(1)		&5.95		&1.5		\\ \hline
%=========================
Ar			&4			&-5.1182368(2)			&-5.1182376(4)		&7.53		&---		\\
Ne			&			&-3.464174(8)			&-3.464229(13)		&4.67		&1.4		\\
{\tiny $1\over 2$}-Ne	&			&-2.64356(3)			&-2.64383(4)		&3.74		&1.8		\\ \hline
%=========================
Ar			&5			&-7.78598(1)			&-7.7862(5)		&4.23		&2.1		\\
Ne			&			&-5.29948(8)			&-5.3037(3)		&2.79		&2.0		\\
{\tiny $1\over 2$}-Ne	&			&-4.0669(1)			&-4.0748(5)		&2.55		&1.5		\\
%=========================
\end{tabular}

\vskip 0.5cm
\caption[Comparison between \DMC\ and \VMC\ numerical results.]
{\footnotesize
Estimates of the ground state energies $E_0$ for noble gases Ar and Ne
and hypothetical lighter particle ${1\over 2}\mbox{-Ne}$ obtained by
using improved diffusion algorithm compared with estimates $E_{\rm T}$
taken from Ref.~\cite{andrei}.  Standard errors in the last digit are
given in parentheses.  Estimates of the relative errors, as described
in text, are given by $Q'$ and $Q''$.  Missing values indicate cases
where the statistical errors are smaller than the errors due to
numerical differentiation.
\par\vspace{5mm}}
\label{tab.compare}
\end{table}

\begin{figure}
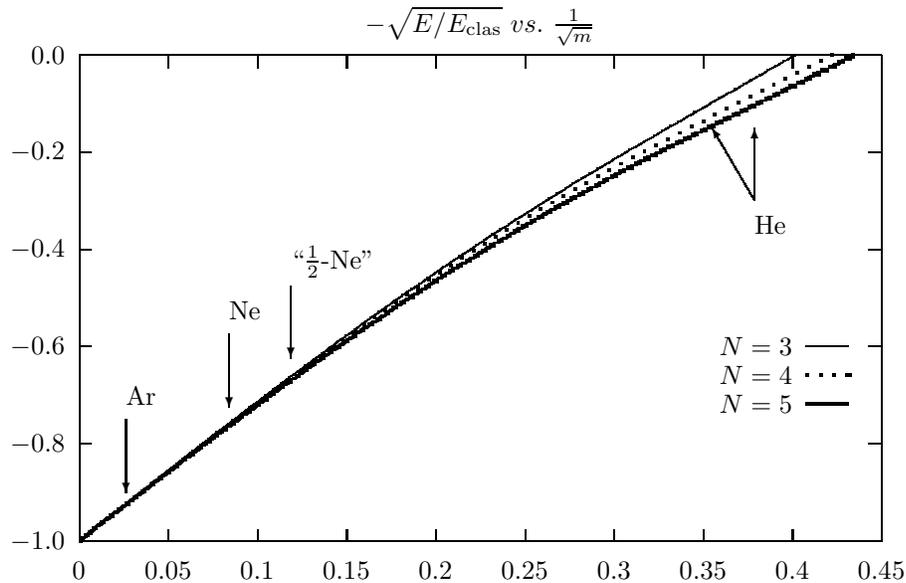

\centerline{\input e}
\vskip 0.5cm
\caption[Critical mass curves.]
{\footnotesize $-\sqrt{E_0/E_{\rm clas}}$ {\it vs.}
de Boer parameter, $1/\sqrt{m}$.}
\label{fig.e}
\end{figure}

\begin{figure}
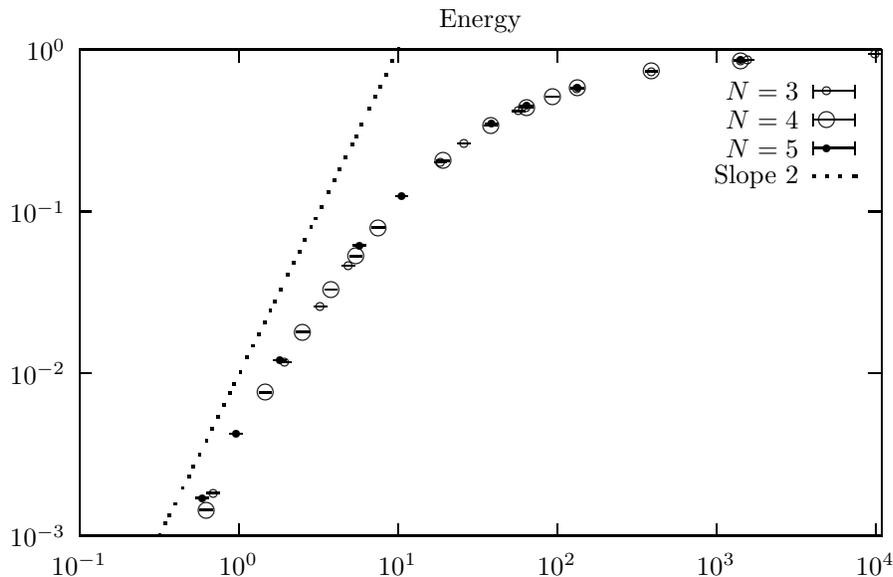

\centerline{\input log.e}
\vskip 0.5cm
\caption[Log-log plot for energies.]
{\footnotesize
Log-log plot of the normalized ground state energy
$E_0/E_{\rm clas}$ {\it vs.} the deviation from the un-binding
transition, $m-m_{\rm c}$.}
\label{fig.log.e}
\end{figure}

Next we compare the behavior of the ground state energy on mass with the
behavior predicted by Eqs.~(\ref{eq.critical}).
Fig.~\ref{fig.e} shows the energy as a function of the de Boer parameter
for clusters of sizes $N = 3,4$ and $5$.  The energy has been normalized
by dividing by the classical ground state energy and we note that the
linear behavior for small de Boer parameter, i.e., large mass, follows
from the harmonic approximation.  Fig.~\ref{fig.log.e} explicitly shows
the data in a double-logarithmic plot of the ground state energy {\it vs}
the deviations from the respective critical points.

\begin{figure}
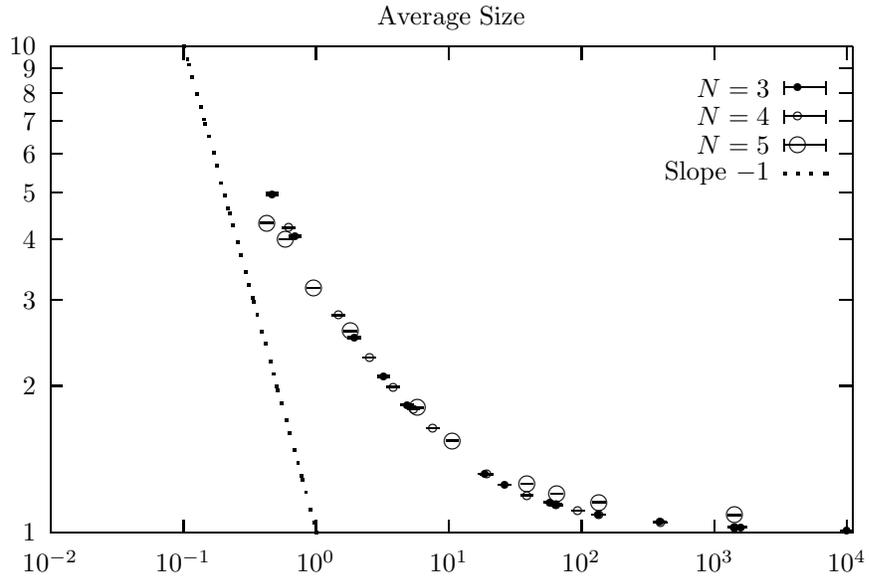

\centerline{\input log.r}
\vskip 0.5cm
\caption[Log-log plot for average size.]
{\footnotesize Log-log plot of the average cluster size  $r$ as defined
in the text {\it vs.} the deviation from the un-binding
transition, $m-m_{\rm c}$.}
\label{fig.log.r}
\end{figure}

\begin{figure}
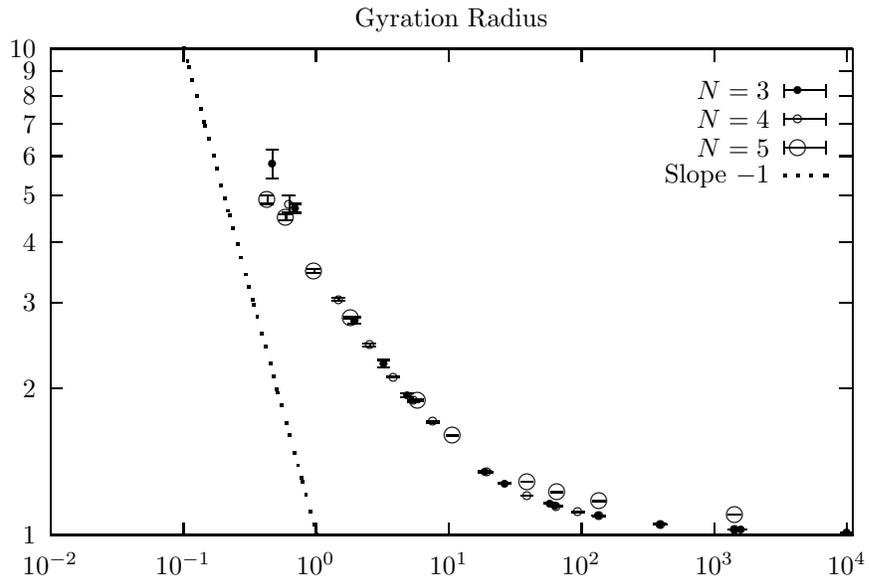

\centerline{\input log.R}
\vskip 0.5cm
\caption[Log-log plot for gyration radius.]
{\footnotesize Log-log plot of the gyration radius $R$
{\it vs.} the deviation from the un-binding
transition, $m-m_{\rm c}$.}
\label{fig.log.R}
\end{figure}

The average size of the clusters was expressed in terms of $r_{ij}$,
the distance between atoms $i$ and $j$, using the following two
definitions:  {\it (1)} $r= 2\langle r_{ij} \rangle/N(N-1)$ {\it (2)}
$R=(2\langle r_{ij}^2 \rangle/N(N-1))^{1/2}$. Double logarithmic plots
of the average size {\it vs} the deviations from the respective
critical points are shown in Figs.  \ref{fig.log.r} and
\ref{fig.log.R}.  It should be noted that in contrast to the
\dMC\ estimates of the energy, the estimates of the average size are
biased in the sense that for any operator $A$ that does not commute
with the hamiltonian \dMC\ yields the matrix element $\langle \psi_{\rm
T}|A|\psi_0 \rangle$ rather than the ground state expectation value
$\langle \psi_0|A|\psi_0 \rangle$.  The values shown in the figures
were obtained by linear extrapolation from $\langle \psi_{\rm
T}|A|\psi_{\rm T} \rangle$ and $\langle \psi_{\rm T}|A|\psi_0
\rangle$.  Irregularities in the quality of the trial functions are
presumably responsible for the corresponding irregularities in the
average cluster sizes for the smallest masses in the $N=5$ case.

\section{Conclusions}
\label{Conclusions}

We presented estimates of the ground state energies of van der Waals clusters
in a wide range of masses.  The results show that wave functions introduced in
Ref. \onlinecite{andrei} provide good trial function in the whole range
from the classical limit to the ultra-quantum limit, where the clusters cease
to form a bound state.  Results for the behavior in the vicinity of the
un-binding transition corroborates Eqs.~({eq.critical}).  The agreement for the
case of the energy is very good; for the divergence in the size dependence
the average cluster size of the clusters closest to the transition was not
quite large enough to show the asymptotic ``critical behavior."

\acknowledgements
It is a great pleasure to acknowledge numerous discussions with David
Freeman, Alex Meyerovich, Cyrus Umrigar and Sergei Stepaniants.
This work was supported by
the Office of Naval Research and by NSF Grants Nos. DMR-9214669 and
CHE-9203498.

%\clearpage

\end{document}